\documentclass[preprint,showpacs,amsmath,amssymb,showkeys,amsfonts,preprintnumbers]{revtex4}
\usepackage{graphicx}
\usepackage{dcolumn}
\usepackage{bm}

\begin{document}
\title{Towards a Supersymmetric Generalization of the Schwarzschild Black Hole}
\author{J.~C.~L\'{o}pez-Dom\'{\i}nguez}
 \email{jlopez@fisica.ugto.mx}
\author{O.~Obreg\'{o}n}
 \email{octavio@fisica.ugto.mx}
\author{S.~Zacar\'{\i}as}
 \email{szacarias@fisica.ugto.mx}
\affiliation{Departamento de F\'{\i}sica, Divisi\'on de Ciencias e
Ingenier\'ias, Campus Le\'on, Universidad de Guanajuato, P.O. Box
E-143, Le\'{o}n, Guanajuato, M\'{e}xico.}
\date{\today}
\begin{abstract}
The Wheeler-DeWitt (WDW) equation for the Kantowski-Sachs model can
also be understood as the WDW-equation corresponding to the
Schwarzschild black hole due to the well known diffeomorphism
between these two metrics. The WDW-equation and its solutions are
``ignorant'' of the coordinate patch one is using, only by imposing
coordinate conditions we can differentiate between cosmological and
black hole models. At that point, the foliation parameter $t$ or $r$
will appear in the solution of interest. In this work we
supersymmetrize this WDW-equation obtaining an extra term in the
potential with two posible signs. The WKB method is then applied,
given rise to two classical equations. It is shown that the event
horizon can never be reached because, very near to it the extra term
in the potential, for each one of the equations, is more relevant
than the one that corresponds to Schwarzschild. One can then study
the asymptotic cases in which one of the two terms in the
Hamiltonian dominates the behavior. One of them corresponds to the
usual Schwarzschild black hole. We will study here the other two
asymptotic regions; they provide three solutions. All of them have a
singularity in $r=0$ and depending on an integration constant $C$
they can also present a singularity in $r=C^2$. Neither of these
solutions have a Newtonian limit. The black hole solution we study
is analyzed between the singularity $r=C^2$ and a maximum radius
$r_m$. We find an associated mass, considering the related
cosmological solution inside $r=C^2$, and based on the holographic
principle an entropy can be assigned to this asymptotic solution.
\end{abstract}

.


\maketitle

\section{Introduction}
Black hole physics has been extensively studied in the literature.
It is useless to try to address the many interesting aspects, even
those related with a single topic. A very rich discussion exists in
the literature in one of these topics, namely that concerning black
holes (event) horizons. One can begin by mentioning the fact that in
general relativity \cite{libros} for stationary vacuum solutions to
the Einstein field equations event horizons arise. Moreover,
classical collapse of astrophysical objects results in (future)
event horizons \cite{harrison}. The existence of an event horizon
means that one has an inaccesible region, and therefore an external
observer must then consider hidden states; pure states become
density matrices. So that, seen from outside, the evolution results
non--unitary, there is information loss. This is one of those things
one has to live with, if one accepts the usual Carter--Penrose
diagram. Modifications to this diagram have been proposed based on
different theoretical frameworks and models all hinting to a more
subtle history for collapse \cite{Visser}. In classical numerical
relativity calculations, event horizons are almost impossible to
find with any certainty. Other definitions of horizons like local or
quasi--local are used to be able to perform calculations that make
sense \cite{numericos}. It has also been claimed by several authors
(see \cite{joshi} and works cited therein) that there is a variety
of physically realistic stellar collapse scenarios in which an event
horizon does not in fact forms, so that the singularity remains
exposed. Moreover, eventhough astronomers will recognize that what
they have observed so far is compatible with the Schwarzschild or
Kerr metrics \cite{Melia}, they will also argue that one can not
unambiguously conclude that the dark objects they observe are black
holes in the sense of general relativity. There is an increasing
consensus, or at least suspicion, within the general relativity
community that event horizons are simply the wrong thing to be
looking at. Other possible definitions of horizons have then been
proposed; apparent \cite{Visser:1995cc}, dynamical \cite{ashtekar},
trapping \cite{Hayward} horizons that make more physical sense. Very
powerful and sophisticated methods have been developed since the
birth of general relativity searching for solutions to its field
equations. For a long time it has been known that changing the
structure of spacetime (i.e. interchanging the coordinates $t
\leftrightarrow r$), changes a static solution for a cosmological
one and viceversa. The best known case is the Schwarzschild metric
that under this particular diffeomorphism transforms into the
Kantowski-Sachs metric \cite{Kantowski:1966te}. This interchange of
variables has been recently proposed as a method to generate new
cosmological models from stationary axisymmetric solutions
\cite{Obregon:2004tp}. In string theory, it has been suggested that
by interchanging $r \leftrightarrow it$ we can get time dependent
solutions also from static and stationary solutions. In this way we
may relate Dp-branes solutions to S-brane solutions, i.e.
time-dependent backgrounds of the theory \cite{Quevedo}. On the
other hand, there are proposals to obtain directly S-brane solutions
\cite{Leblond}; thus, if cosmological solutions (i.e. S-branes) can
be generated from stationary ones (i.e. Dp-branes), this procedure
also works the other way around.

In particular, for a Schwarzschild black hole, $\rm Kucha\check{r}$
\cite{Kuchar:1994zk} has shown how to reconstruct the curvature
coordinates $T$ and $R$ (or the Kruskal coordinate $U$ and $V$) from
spherically symmetric initial data. His formalism makes possible a
discussion of the action of space-time diffeomorphism on the quantum
geometry. A particular interesting example is the interchange of the
curvature coordinates $T$ and $R$. This choice of coordinates
interchanges the static and dynamical regions for the Kruskal
diagram transforming the Kantowski-Sachs cosmological metric into
the Schwarzschild metric. This relation was taken into account
suggesting a canonical approach based on a foliation in the
parameter $r$, by this means a Hamiltonian formalism is developed.
This kind of approach was used to find quantum black hole states
\cite{Cavaglia:1994yc} and a generalization to a noncommutative
minisuperspace provides a particular model towards the understanding
of noncommutative quantum black holes \cite{LopezDominguez:2006wd}.
The WDW-equation solutions are ``ignorant" of the coordinate patch
one is using and only when we impose coordinate conditions will
there be any difference between cosmological models and black hole
solutions. Only at that point will the foliation parameter $t$ and
$r$, respectively, appear in the solution of interest.

Several approaches have been suggested to supersymmetrize the
WDW-equation for cosmological models; the first model proposed
\cite{Macias:1987ir} was based on the fact that shortly after the
invention of supergravity \cite{Freedman:1976xh} it was shown
\cite{Teitelboim:1977hc}, that this theory provides a natural
classical square root of gravity. By this means a method for finding
square root equations and their corresponding Hamiltonians in
quantum cosmology was proposed in \cite{Macias:1987ir}, that is the
study of supersymmetric quantum cosmology. Later, a superfield
formulation was introduced, by means of which it is possible to
obtain, in a direct manner, the corresponding fermionic partners and
also being able to incorporate matter in a simpler way \cite{tkach}.
A third method allows to define a ``square root'' of the potential,
in the minisuperspace, of the cosmological model of interest and
consequently operators which square results in the Hamiltonian
\cite{grahamyotros}, other related proposals have been studied
\cite{obregon}. Being the minisuperspace variables in the WDW-
equation, and consequently the corresponding wave functions,
``ignorant'' of the coordinate patch one is using, one imposes
coordinate conditions which then produce the difference between
cosmological solutions and black hole models.

In this work a WDW-equation for a Schwarzschild black hole, is
considered, \cite{Kantowski:1966te,Cavaglia:1994yc}. It is
explicitly shown that, by means of the WKB method, one gets the well
known Schwarzschild solution. Making use of the third method
mentioned above \cite{grahamyotros}, a quantum supersymmetric
Kantowski-Sachs model, and consequently its corresponding
supersymmetric quantum black hole model is found. We get operators
which square provides two Hamiltonians that generalize the
WDW-equation. A simple WKB approach is applied to these Hamiltonians
leading to two classical equations, having each one, two asymptotic
regions that can be analytically obtained. The Schwarzschild black
hole is one of these asymptotic solutions in both cases. However, in
general, its horizon can never be reached because when $2m/r$ is
very near to one the other two asymptotic regions, of each one of
the equations correspondingly are the valid ones. The analytic
solutions can be found for both of them, they are singular at $r=0$;
depending on an integration constant $C$, another singularity
appears at $r=C^2$. In these asymptotic regions, neither of the
solutions have a Newtonian limit. Eventhough, these asymptotic
regions are consequence of supersymmetry, it results of interest to
analyze their corresponding classical solutions to understand the
behavior of the general classical solution in these asymptotic
regions that drastically differ from the Schwarzschild one. It has
been shown \cite{Gibbons}, by solving the Dirac equation,
particularly, in the Schwarzschild and Kerr backgrounds that the
spinors blow up at the horizon. In our supersymmetric black hole
model the fermionic degrees of freedom are intrinsic elements of the
theory and they do not allow the presence of the Schwarzschild
horizon. Solutions of supergravity theories played a crucial role in
important developments in string black hole physics, AdS/CFT and
others. It is well known that massive neutral particles cannot be
associated with supersymmetric BPS-states, the most simple
spherically symmetric solution that admits Killing spinors to
satisfy the constrains that define BPS-states is the
Reinner-Nordstr\"{o}m black hole with $M=\left|Q\right|$\cite{bh}. The
classical (and quantum) supersymmetric Schwarzschild black hole
model we propose is based, as mentioned above, in supersymmetrizing
the WDW-equation associated with the standard Schwarzschild black
hole. This procedure provides a modified (SUSY quantum)Hamiltonian
and its corresponding classical equations that, {\it in this sense}
define a supersymmetric generalization of the Schwarzschild black
hole. Our proposal seems to provide a starting point to understand
and construct a first model of a classical supersymmetric
Schwarzschild black hole.

If we would apply whole supergravity ($\mathcal{N}=1$) to the
Kantowski-Sachs-Schwarzschild model, instead of directly
supersymetrizing its WDW-equation, it is to be expected to get and
equivalent Hamiltonian. As already outlined, in this work we will
analyze, in the context of the minisuperspace approximation, the
generalized supersymmetric WDW-equation for the oldest and most well
known black hole that was discovered by Schwarzschild. We will first
review, in section II, the WDW-equation for a Schwarzschild black
hole, taking advantage of its diffeomorphism with the
Kantowski-Sachs model \cite{Kantowski:1966te,Cavaglia:1994yc}, and
will use the WKB method to obtain the corresponding well known
classical solution. In section III we choose the third and simplest
of the three approaches, outlined above, to supersymmetrize the
WDW-equation and apply it to define the supersymmetric model for the
Schwarzschild black hole. The same kind of result would essentially
be obtained by using any of the other two approaches (in the case of
the first of them, based directly in supergravity $\mathcal{N}=1$,
the fermionic partners are the gravitinos). In this way the
corresponding (super) Hamiltonian is obtained. In section IV, the
classical analysis, a WKB approach is performed for the only two
diagonal components of the (super) Hamiltonian, and the equations
corresponding to the asymptotic regions are analytically solved. As
already stated, the standard Schwarzschild metric is the solution to
one of them and there are other two asymptotic regions corresponding
to each one of the two Hamiltonians. One has two solutions and the
other, only one, they have a singularity at $r=0$, and can present
also another singularity at $r=C^2$, $(C={\rm constant})$. The
Schwarzschild horizon can never be reached because when $2m/r$ is
very near to one, one must consider the other asymptotic region, in
each case, and their corresponding solutions. Eventhough, in the
framework of our proposal, the Schwarzschild solution and the
``supersymmetric'' solutions are asymptotic solutions, it seems of
interest to study the behavior of the last ones; in section V we
analyze one of them that has the two singularities at $r=0$ and at
$r=C^2$, find its associated mass in this asymptotic region, and by
means of the cosmological model inside $r=C^2$ and making use of the
holographic principle, we are able to propose an entropy related to
this solution and show its relation with the mass. Section VI is
devoted to discussion and conclusions.


\section{WDW equation for
Schwarzschild$\mathbf{\rightarrow}$Kantowski-Sachs\\
metrics and the classical limit.}\label{seccion2}

Let us begin by reviewing the relationship between the cosmological
Kantowski-Sachs metric and the Schwarzschild metric
\cite{Kantowski:1966te,Cavaglia:1994yc}. The Schwarzschild solution
can be written as
\begin{align}
ds^{2}=-\left(1-\frac{2m}{r}\right)dt^{2}+\left( 1-\frac{2m}
{r}\right)^{-1}dr^{2}+r^{2}\left(d\theta^{2}+ \sin^{2}\theta
d\varphi^{2}\right).\label{shcwar}
\end{align}
For the case $r<2m$, the $g_{tt}$ and $g_{rr}$ components of the
metric change in sign and $\partial_{t}$ becomes a space-like
vector, and $\partial_r$ becomes a time--like vector. If we make the
coordinate transformation $t\leftrightarrow r$, we find
\begin{align}
ds^{2}=-\left(\frac{2m}{t}-1\right)^{-1}dt^{2}+\left(\frac{2m}
{t}-1\right)dr^{2}+t^{2}\left(d\theta^{2}+\sin^{2}\theta
d\varphi^{2}\right)\label{KSsch}.
\end{align}
On the other hand, the parametrization by Misner \cite{misner}
appropriate for the Kantowski-Sachs and Schwarzschild metrics is
\begin{align}
ds^{2}=-N^{2}dt^{2}+e^{2\sqrt{3}\beta}dr^{2}+e^{-2\sqrt{3}\beta}
e^{-2\sqrt{3}\Omega}\left(d\theta^{2}+\sin^{2}\theta
d\varphi^{2}\right). \label{KSmetric}
\end{align}
The corresponding WDW-equation for the Kantowski-Sachs metric,
results in
\begin{align}
\left[  -\frac{\partial^{2}}{\partial\Omega^{2}}+\frac{\partial^{2}}%
{\partial\beta^{2}}+48e^{ -2\sqrt{3}\Omega}\right]
\psi(\Omega,\beta)=0.\label{ks}
\end{align}
The solution to this equation was given by Misner \cite{misner}.

Based on the diffeomorphism between the Kanstowski-Sachs and the
Schwarzschild metrics, the WDW-equation (\ref{ks}) has been applied
to find a quantized version of a Schwarzschild black hole
\cite{Cavaglia:1994yc}. As mentioned, our objective is to find the
(super)WDW-equation corresponding to (\ref{ks}) that can be traduced
in a (super)Hamiltonian for a Schwarzschild black hole and will
concentrate our study to its classical solutions, {\it i.e.} the
supersymmetric generalization of the Schwarzschild black hole. In
order to obtain that, first we show how to get the solutions
(\ref{shcwar}) and (\ref{KSsch}) by applying the WKB method to the
WDW-equation (\ref{ks}). As well known, we assume that the wave
function has the form
\begin{gather}
\psi=e^{i[S_{1}(\Omega)+S_{2}(\beta)]}~.\label{functionS}
\end{gather}
The usual procedure results in the Einstein-Hamilton-Jacobi equation
\begin{gather}
-\left(\frac{dS_{1}(\Omega)}{d\Omega}\right)^2+
\left(\frac{dS_{2}(\beta)}{d\beta}\right)^2-
48e^{-2\sqrt{3}\Omega}=0,
\end{gather}
one identifies
\begin{gather}
\frac{dS_{1}(\Omega)}{d\Omega}\rightarrow\Pi_{\Omega}~~~~ {\rm
and}~~~~ \frac{dS_{2}(\beta)}{d\beta}\rightarrow\Pi_{\beta},
\end{gather}
where
\begin{gather}
\Pi_{\Omega}=-\frac{12}{N}e^{-\sqrt{3}\beta-2\sqrt{3}\Omega}\dot{\Omega}
~~~~{\rm and}~~~~
\Pi_{\beta}=\frac{12}{N}e^{-\sqrt{3}\beta-2\sqrt{3}\Omega}\dot{\beta},
\end{gather}
then the classical equation to be solved is
\begin{gather}
\frac{3}{N^{2}}\left(\dot{\Omega}^{2}-\dot{\beta}^{2}\right)+
e^{2\sqrt{3}\Omega+2\sqrt{3}\beta}=0.\label{ecclasica}
\end{gather}
Making use of the Misner parametrization (\ref{KSmetric}), taking
$e^{-2\sqrt{3}\Omega-2\sqrt{3}\beta}=t^2$, and identifying
$N^2=e^{-2\sqrt{3}\beta}$, we get the equation
\begin{gather}
e^{-2\sqrt{3}\Omega}\left(1+2\sqrt{3}t\dot{\Omega}\right)-t^2=0,
\label{ecclasica2}
\end{gather}
which solution
\begin{gather}
e^{-2\sqrt{3}\Omega}=2mt-t^2,\label{solucionKS}
\end{gather}
with $m=$ constant, brings us back to the metric (\ref{KSsch}), and
due to the diffeomorphism between the solutions (\ref{shcwar}) and
(\ref{KSsch}), choosing the parameter $r$ instead of $t$, we get the
Schwarzschild solution (\ref{shcwar}) as well. So, as it should be,
the WKB method applied to the WDW-equation (\ref{ks}) gives the
classical equation (\ref{ecclasica2}) and its solution
(\ref{solucionKS}) is the same as that of the classical Einstein
equations.

\section{A supersymmetric WDW equation for the Misner parametrization
of the Kantowski--Sachs--Schwarzschild metrics.}\label{seccion3}

In order to generalize the WDW-equation (\ref{ks}) to its
supersymmetric version the third method outlined in the introduction
will be used \cite{grahamyotros,Lidsey,kiefer}. The Hamiltonian for
the homogeneous models can in general be written as
\begin{align}
2H_0=G^{\mu\nu}\Pi_\mu\Pi_\nu+U(q),
\end{align}
where $G^{\mu\nu}$ is the metric in the minisuperspace. It's
possible to find a function $\phi$ such that
\begin{gather}
G^{\mu \nu}\frac{\partial \phi}{\partial q^{\mu}}\frac{\partial\phi}
{\partial q^{\nu}}=U(q).\label{gmunu}
\end{gather}
Thus, the minisuperspace Hamiltonian is written in the form
\begin{gather}
H=\frac{1}{2}\left[Q\overline{Q}+\overline{Q}Q\right] =
H_0+\frac{1}{2}\frac{\partial^2\phi}{\partial q^\mu\partial
q^\nu}\left[\bar{\theta}^\mu,\theta^\nu\right],\label{superhamiltonian}
\end{gather}
with the non-Hermitian supercharges
\begin{align}
Q=\theta^{\mu}\left(\Pi_{\mu}+ i\frac{\partial \phi}{\partial
q^{\mu}}\right),~~
\overline{Q}=\overline{\theta}^{\mu}\left(\Pi_{\mu}
-i\frac{\partial\phi}{\partial q^{\mu}}\right),\label{supc}
\end{align}
where $\theta^{\nu}$ and $\bar{\theta}^{\nu}$ satisfy the spinor
algebra
\begin{align}
\left\{\bar{\theta}^{\mu},\bar{\theta}^{\nu}\right\}=0,~~~~
\left\{\theta^{\mu},\theta^{\nu}\right\}=0,~~~~
\left\{\bar{\theta}^{\mu},\theta^{\nu}\right\}=G^{\mu\nu}.
\label{algebragrass}
\end{align}
For our model (\ref{ks}) $U=-48e^{-2\sqrt{3}\Omega}$ and the
Hamilton-Jacobi equation is then
\begin{align}
-\left(\frac{\partial\phi}{\partial\Omega}\right)^2+\left(\frac{\partial\phi}
{\partial\beta}\right)^2=-48e^{-2\sqrt{3}\Omega},\label{ecHJ}
\end{align}
a solution is
\begin{align}
\phi=-4e^{\sqrt{3}\Omega},
\end{align}
then according to (\ref{supc}) and (\ref{algebragrass}) the
supercharges are given by
\begin{gather}
Q =\theta^{\Omega}(\Pi_{\Omega}+i4\sqrt{3}e^{-\sqrt{3}\Omega})+
\theta^{\beta}\Pi_{\beta}, \nonumber\\
\overline{Q}=\overline{\theta}^{\Omega}(\Pi_{\Omega}-
i4\sqrt{3}e^{-\sqrt{3}\Omega})+\overline{\theta}^{\beta}\Pi_{\beta},
\label{QbarQ}
\end{gather}
where $\Omega$ and $\beta$ are the minisuperspace coordinates.

To obtain the supersymmetric Hamiltonian operator it is necessary to
find appropriate representations for the bosonic variables and the
fermionic ones $\theta^{\Omega}$, $\bar{\theta}^{\Omega}$,
$\theta^{\beta}$ and $\bar{\theta}^{\beta}$. The momenta will be the
usual differential operators
$\Pi_\Omega\rightarrow-i\dfrac{\partial}{\partial\Omega}$,
$\Pi_\beta\rightarrow-i\dfrac{\partial}{\partial\beta}$ and to
realize the fermionic variables algebra (\ref{algebragrass}) we will
represent them as matrices, in the following manner
\begin{align}
&2\widehat{\theta}^{\Omega}=\gamma^{1}-i\gamma^{2},~~~~
2\widehat{\bar{\theta}}^{\Omega}=\gamma^{1}+i\gamma^{2},\nonumber\\
&2\widehat{\theta}^{\beta}=\gamma^{0}+\gamma^{3},~~~~~
2\widehat{\bar{\theta}}^{\beta}=\gamma^{0}-\gamma^{3},\label{gammas}
\end{align}
for the $\gamma$--matrices we will use the representation
\begin{align}
&\gamma^{0}=\left(\begin{array}{cccc}0 & 0 & 0 & -i \\
0 & 0 & i & 0 \\0 &- i & 0 & 0 \\i & 0 & 0 & 0\end{array}\right),
~~~ \gamma^{1}=\left(\begin{array}{cccc}i & 0 & 0 & 0 \\0 & -i & 0 &
0\\0 &  0 & i & 0 \\0 & 0 & 0 & -i \end{array}\right),\nonumber\\
&\gamma^{2}=\left(\begin{array}{cccc}0 & 0 & 0 & i \\
0 & 0 & -i & 0 \\0 & -i & 0 & 0 \\i & 0 & 0 &0\end{array}\right),~~~
\gamma^{3}=\left(\begin{array}{cccc} 0& -i & 0 & 0 \\-i & 0 & 0 & 0
\\0 &  0 & 0 & -i \\0 & 0 & -i & 0 \end{array}\right).\label{gammas2}
\end{align}
Making use of these operators representation, the $\widehat{Q}$ and
$\widehat{\overline{Q}}$ operators can be constructed from
(\ref{QbarQ}) and with them, a diagonal Hamiltonian operator
$\widehat{H}$ is obtained, since the first and third as well as the
second and fourth operators are equal, the wave function has only
two components. The usual WDW-equation (\ref{ks}) is by these means
generalized to two quantum equations for the
Kantowski-Sachs-Schwarzschild minisuperspace, namely
\begin{gather}
-\frac{\partial^2}{\partial\Omega^{2}}\psi_{\pm}+
\frac{\partial^2}{\partial\beta^{2}}\psi_{\pm}+
12e^{-2\sqrt{3}\Omega}\left(4\pm
e^{\sqrt{3}\Omega}\right)\psi_{\pm}=0, \label{sol}
\end{gather}
where $\psi_{+}$ and $\psi_{-}$ correspond to the wave function
associated with the $(+)$ and the $(-)$ signs in the potential.

Note that both quantum equations differ from the usual WDW-equation
(\ref{ks}) by the same extra term but with different sign. This is
similar to what happens in standard supersymmetric quantum mechanics
where also extra terms arise in the potential. Eventhough, the last
term in (\ref{sol}), is expected to be relevant only when the
supersymmetric contribution is larger or of the order of the usual
one, it is of interest to study the behavior of these other
asymptotic solutions that considerable differ from the Schwarzschild
solution, as will be shown.

It is clear that to certain linear combinations of the fermionic
operators $\widehat{{\theta}}^{\Omega}$,
$\widehat{{\theta}}^{\beta}$, $\widehat{{\bar{\theta}}}^{\Omega}$
and $\widehat{{\bar{\theta}}}^{\beta}$ one can associate the
eigenvectors
\begin{align}
\left(\begin{array}{c} 1\\ 0\\ 0\\0
\end{array}\right),
\left(\begin{array}{c} 0\\ 1\\ 0\\0
\end{array}\right),
\left(\begin{array}{c} 0\\ 0\\ 1\\0
\end{array}\right)~~{\rm and}~~
\left(\begin{array}{c} 0\\ 0\\ 0\\1
\end{array}\right).\label{eigenvec}
\end{align}
So, for example, the first eigenvector is an eigenstate with
eigenvalue $+1$ corresponding to one of these particular
combinations, having then that the eigenvectors (\ref{eigenvec}) are
linear combinations of the eigenstates of the operators
(\ref{gammas},\ref{gammas2}), through an appropriate rotation one
could associate the above four eigenstates directly with the
fermionic operators $\widehat{{\theta}}^{\mu}$ and
$\widehat{{\bar{\theta}}}^{\mu}$ \cite{Macias:1987ir,obregon}, and
by this means identify their contribution to wave function
components.

\section{The classical limit of the Supersymmetric region}

Following the same procedure, to apply the WKB method presented in
section \ref{seccion2}, one obtains two classical equations
corresponding to (\ref{sol}), that written together give
\begin{align}
\frac{12}{N^{2}}\left(\dot{\Omega}^{2}-\dot{\beta}^{2}\right)+
e^{2\sqrt{3}\Omega+2\sqrt{3}\beta}\left(4\pm
e^{\sqrt{3}\Omega}\right)=0.\label{susywdw}
\end{align}
Making use of the Misner parametrization (\ref{KSmetric}), as in
section \ref{seccion2} and using
$e^{-2\sqrt{3}\Omega-2\sqrt{3}\beta}=t^2$ and
$N^2=e^{-2\sqrt{3}\beta}$ one has
\begin{align}
4e^{-2\sqrt{3}\Omega}\left(1+2t\sqrt{3}\dot{\Omega}\right)-t^2
\left(4\pm e^{\sqrt{3}\Omega}\right)=0.\label{clasusy}
\end{align}
Both of the classical equations (\ref{clasusy}) present two
asymptotic regions for $4\gg e^{\sqrt{3}\Omega}$ and another for
$4\ll e^{\sqrt{3}\Omega}$. By interchanging $t$ by $r$, as in
section \ref{seccion2}, the first limit evidently gives the
Schwarzschild solution (\ref{shcwar}), (\ref{ecclasica2}) and
(\ref{solucionKS}). In the other region the extra term in the
potential (\ref{clasusy}) dominates and one has two equations. We
will see that the corresponding solutions present a singularity at
$r=0$ and depending on the sign of the integration constant $C$,
another singularity could also be present at $r=C^2$ in both cases.
In these ``supersymmetric'' dominated regions, neither of the
solutions can be related with the Newtonian limit. Already, at this
stage, equation (\ref{clasusy}) tells us that the Schwarzschild
horizon can never be reached.

As we can deduce from (\ref{solucionKS}), the Schwarzschild case,
changing $t$ by $r$, gives
\begin{align}
e^{\sqrt{3}\Omega}=\frac{1}{r\left(1-\frac{2m}{r}\right)^{1/2}}
\label{constrain}.
\end{align}
Independently of the value of $r$, for $2m/r$ very near to one,
$e^{\sqrt{3}\Omega}$ can be very large, $e^{\sqrt{3}\Omega}\gg 4$
and according to (\ref{clasusy}) we are then in the other asymptotic
regions, not the one corresponding to Schwarzschild and these have
different solutions.

The solutions for these asymptotic regions $\left({\rm when}~~
e^{\sqrt{3}\Omega}\gg 4\right)$, are:
\begin{align}
e^{-\sqrt{3}\Omega}=\left(\frac{3}{4}\right)^{1/3}r^{2/3}
\left(\pm1+\frac{C}{\sqrt{r}}\right)^{1/3},\label{solposneg}
\end{align}
where $C$ is a constant and taking into account the Misner
parametrization (\ref{KSmetric}) and appropriately the coefficients
of $dt^2$ and $dr^2$ we get
\begin{align}
ds^{2}=&-\left(\frac{3}{4}\right)^{2/3}r^{-2/3}
\left(\pm1+\frac{C}{\sqrt{r}}\right)^{2/3}dt^{2}+
\left(\frac{4}{3}\right)^{2/3}r^{2/3}\left(\pm1+\frac{C}{\sqrt{r}}\right)^{-2/3}
dr^{2}\nonumber\\
&+r^{2} \left(d\theta^{2}+\sin^{2}\theta
d\phi^{2}\right).\label{solucionposneg}
\end{align}
As stated, these solutions are valid only in the asymptotic region
$e^{\sqrt{3}\Omega}\gg 4$. Taking the solution with the positive
sign, in the one of the parenthesis, one should consider two cases
$C>0$ and $C<0$. In the first of them, solution 1, there is a
singularity only at $r=0$. In the second case there is also a second
singularity at $r=C^2$, solution 2. For the solution with the
negative sign the constant $C$ can be negative or positive. The
first possibility would give a negative value for
$e^{-\sqrt{3}\Omega}$, then this is not a valid case. So, we will
consider only values $C>0$. The singularities for this case will be
at $r=0$ and at $r=C^2$, solution 3. In order to exhibit the
singularities, the Kretschmann invariant is calculated for the
solutions (\ref{solucionposneg}), it is given by
\begin{align}
&K_{\pm}=\frac{1}{864\left(\pm
1+\frac{C}{\sqrt{r}}\right)^{8/3}~r^{22/3}}\times\nonumber\\
&\times\left[
\begin{array}{c}
3888~6^{1/3}C^4-216~6^{1/3}C^3r^{1/2}\left[\mp63+8~6^{1/3}\left(\pm1+\frac{C}
{\sqrt{r}}\right)^{1/3}r^{2/3}\right] \\
+32r^2\left[71~6^{1/3}\mp54~6^{2/3}r^{2/3}\left(\pm1+\frac{C}{\sqrt{r}}\right)
^{1/3}+108r^{4/3}\left(\pm1+\frac{C}{\sqrt{r}}\right)^{2/3}\right]\\
\pm24Cr^{3/2}\left[431~6^{1/3}\mp216~6^{2/3}r^{2/3}\left(\pm1+\frac{C}
{\sqrt{r}}\right)^{1/3}+288r^{4/3}\left(\pm1+\frac{C}{\sqrt{r}}\right)
^{2/3}\right] \\
+27C^2\left[659~6^{1/3}r\mp192~6^{2/3}r^{5/3}\left(\pm1+\frac{C}{\sqrt{r}}
\right)^{1/3}+128r^{7/3}\left(\pm1+\frac{C}{\sqrt{r}}\right)^{2/3}
\right]\end{array}\right].\label{kretsch}
\end{align}
From (\ref{kretsch}) we can read the above discussed singularities
at the origin $r=0$ and at the radius $r=C^2$.

The SUSY-WDW-equation (\ref{sol}) and their corresponding classical
equations (\ref{susywdw}) have in their own structure the
information of certain states related with our fermionic variables
(``gravitinos'' if we would have used the supergravity
$\mathcal{N}=1$ method), they manifest themselves in the potential.
As well known, by solving the Dirac equation in the Schwarzschild
and Kerr backgrounds the spinors blow up at the horizon
\cite{Gibbons}. It is then reasonable to expect that a model where
the fermionic fields are intrinsically incorporated do not allow the
existence of a horizon, as we have shown. It is to be noted that our
(quantum)supersymmetric approach removes, in the WKB limit
(\ref{solposneg}), in this SUSY asymptotic region, the horizon but
not the singularity at $r=0$ that remains in all the three
solutions, in fact, in two of them another singularity arises in
$r=C^2$.

\section{Analysis of solution 2, an associated mass and a
possible relation with the entropy}

Lets us consider
\begin{align}
ds^2=-Fdt^2+F^{-1}dr^2+r^2\left(d\theta^2+\sin^2{\theta}d\phi^2\right),
\label{metricaentropia}
\end{align}
with $F=\left(\dfrac{3}{4}\right)^{2/3}r^{-2/3}
\left(1-\dfrac{C}{\sqrt{r}}\right)^{2/3}$ and $C>0$. The expression
given by equation (\ref{solposneg}) must be positive. This imposes
the restriction $\sqrt{r}>C$. Also according with equation
(\ref{clasusy}) solution (\ref{metricaentropia}) is valid for
$e^{\sqrt{3}\Omega}\gg4$. This implies (\ref{solposneg}) that there
is a maximum radius $r_m > C^2$ (in Planck lengths) where this
solution is valid. Due to the fact that for large $r$ the Minkowski
metric is not a limit of (\ref{metricaentropia}), one can consider
this metric at $r_m$ as the background metric. With this assumption
we can follow a well known proposal \cite{deser} to associate a mass
to our metric. The mass formula can be expressed as
\begin{align}
M=-\frac{1}{2G_N}\frac{\left|\tilde{g}_{tt}\right|^{1/2}}{\left|\tilde{g}_{rr}\right|^{3/2}}~
r\left(g_{rr}-\tilde{g}_{rr}\right),
\end{align}
where $\tilde{g}_{tt}$ and $\tilde{g}_{rr}$ correspond to the
background metric, in our case $\widetilde{F}$ and
$\widetilde{F}^{-1}$, which for an enough large $r_m$ results in
\begin{align}
\widetilde{F}^{-1}=\left(\frac{4}{3}\right)^{2/3}~r_m^{2/3},
\end{align}
for $r=\alpha r_m$, $\alpha\ll 1$, then
\begin{align}
F^{-1}=\widetilde{F}^{-1}\alpha^{2/3}\left(1+\frac{2}{3}
\frac{C}{\alpha^{1/2}~r_m^{1/2}}\right),
\end{align}
and
$M=-\dfrac{1}{2G_N}\widetilde{F}^2~r\left(F^{-1}-\widetilde{F}^{-1}\right)$,
which results in
\begin{align}
M=\left(\frac{3}{4}\right)^{2/3}\frac{1}{2G_N}
\left[\alpha\left(1-\alpha^{2/3}\right)r_m^{1/3}-
\frac{2}{3}\alpha^{7/6}Cr_m^{-1/6}\right].\label{masa}
\end{align}

As $r_m$ is proportional to $C^2$, we have that
\begin{align}
M\sim C^{2/3}.\label{masa2}
\end{align}
This relation between the constant of integration $C$ and the mass
was to be expected because
$F=\left(\dfrac{3}{4}\right)^{2/3}\left(\dfrac{1}{r}-
\dfrac{C}{r^{3/2}}\right)^{2/3}$, so $C^{2/3}$ would be related with
$r$, in the same manner as the mass appears in the usual static
metric $\sim \dfrac{M}{r}$.

The SUSY-WDW-equation (\ref{sol}) provides in the asymptotic region
$e^{\sqrt{3}\Omega}\gg 4$, the black hole solution
(\ref{metricaentropia}). This has an associated cosmological model
inside $r=C^2$ as it happens in the standard bosonic case
(\ref{shcwar},\ref{KSsch}). This cosmological solution can be
expressed as (\ref{metricaentropia}) but now with
\begin{align}
F=\left(\frac{4}{3}\right)^{2/3}t^{2/3}\left(\frac{C}{\sqrt{t}}-
1\right)^{-2/3}.\label{FconT}
\end{align}
On the other hand, the holographic principle tell us, that for a
given volume $V$ the state of maximal entropy is given by the
largest black hole that fits inside $V$, 'tHooft and Susskind
\cite{'tHooft:1993gx} argued that the microscopic entropy associated
with the volume $V$ should not exceed the Bekenstein-Hawking entropy
$S\leq \frac {A}{4G}$ of a certain black hole with horizon area $A$
equal to the surface area of the boundary of $V$. A particular model
to realize this idea was given by Verlinde \cite{Verlinde:2000wg},
he generalized the Cardy formula \cite{Cardy:1986ie} to arbitrary
spacetime dimensions and proposed that a closed universe has a
Casimir contribution to its energy and entropy and that the Casimir
energy is bounded from above by the Bekenstein-Hawking energy. He
found that $S_{BH}=(n-1)\dfrac{V}{4\pi GR}$, where $n$ is the number
of space dimensions, $V$ is the volume of the universe and $R$ its
radius. This $S_{BH}$ was identified with the holographic
Bekenstein-Hawking entropy of a black hole with the size of the
universe.

The largest posible radius of the universe
(\ref{metricaentropia},\ref{FconT}) can not exceed $C^2$. So, the
largest standard black hole fitting $V$ would have also $r=C^2$ and
accordingly the maximum entropy of this universe should be
$S\sim\dfrac{A}{4G}\sim r^2\sim C^4$ (note that we can not assign an
entropy to our black hole solution outside $r=C^2$
(\ref{metricaentropia}) by following the usual procedure
\cite{hawkinggibbons}, because this solution has not, as already
shown, a horizon). Now taking into account $M\sim C^{2/3}$
(\ref{masa},\ref{masa2}), it results that $S\sim C^4\sim M^{6}$. As
mentioned, we have not found the general solutions to the classical
equations (\ref{susywdw},\ref{clasusy}), the above resulting mass
and entropy were obtained in the extreme supersymmetric (or
fermionic) limit. As consequence, this entropy $S\sim M^6$, with $M$
given by (\ref{masa},\ref{masa2}), should be interpreted as a
correction to the usual one emerging in this (extreme) asymptotic
(quantum)supersymmetric regime.

\section{conclusions}
Based on previous works \cite{Kuchar:1994zk,Cavaglia:1994yc}, we
have first shown that the WDW-equation (\ref{ks}) has a classical
limit (\ref{ecclasica2}) which solution is the Schwarzschild metric
(\ref{shcwar}), (or the Kantowski-Sachs cosmological model
(\ref{KSsch})). A supersymmetrization of (\ref{ks}) results in
(\ref{sol}) which classical limit (\ref{clasusy}) shows already
(\ref{constrain}) that the event horizon can never be reached
because very near to it the other term in the potential in
(\ref{clasusy}) dominates. It is known, by solving the Dirac
equation, that the spinors blow up at the horizon of the
Schwarzschild and Kerr metrics \cite{Gibbons}. In our supersymmetry
black hole proposal the fermionic degrees of freedom (the gravitinos
if we would have used the first supergravity $\mathcal{N}=1$
\cite{Macias:1987ir,obregon} method mentioned in this work) are
intrinsic elements of the model and they do not allow the presence
of an event horizon. Each equation (\ref{clasusy}) exhibits two
asymptotic limits, neither of them correspond to a whole exact
solution of this equation. Nevertheless, the Schwarzschild black
hole is the solution to one of these asymptotic regions. We have
studied the other two possible asymptotic regions, depending on the
sign of the extra term in the potential. The three solutions are
given by (\ref{solucionposneg}). They are singular at $r=0$ and
depending on the sign of $C$ they can also be singular at $r=C^2$
(\ref{kretsch}). None of these ``supersymmetric'' dominated
solutions have a Newtonian limit. We then analyzed the second
solution (\ref{metricaentropia}), the Minkowski space-time is not a
background metric of it. However, there is a maximum radius,
$r_m>C^2$, the metric (\ref{metricaentropia}) calculated at $r_m$ is
assumed as the background metric. By this means a mass (\ref{masa})
can be associated to this solution and it results that we can relate
it with our constant of integration $C$; $M\sim C^{2/3}$
(\ref{masa2}). Inside $r=C^2$ one has the cosmological solution
(\ref{metricaentropia},\ref{FconT}) and according with the
holographic principle proposal, the state of maximum entropy should
correspond to the entropy associated with the largest standard black
hole fitting its volume. The maximum radius of this universe is
$r=C^2$, consequently $S\sim C^4$ and by means of $M\sim C^{2/3}$
(\ref{masa2}), it results that $S\sim M^6$. This arises in the
asymptotic supersymmetric (fermionic) region, so the mass
(\ref{masa},\ref{masa2}) and its associate entropy correspond to
this extreme regime and could be understood as a (quantum)SUSY
correction to the standard black hole entropy.

\begin{acknowledgments}
This work was supported in part by CONACyT M\'exico Grant 51306 and
PROMEP projects.
\end{acknowledgments}


\end{document}